%% file: JHEP_longitudinal_improved-final-03.11.2026.tex
\documentclass[a4paper,11pt]{article}
\usepackage{jheppub}
\usepackage[T1]{fontenc}
\usepackage{lmodern}
\usepackage[utf8]{inputenc}
\usepackage{amsmath,amssymb,bm,mathtools}
\usepackage{graphicx}
\graphicspath{{./}{figures/}}
\usepackage[caption=false]{subfig}
\hypersetup{hidelinks}
\usepackage{placeins}
\usepackage{booktabs,array,longtable}

\usepackage{wrapfig}

\usepackage{mathtools}

\providecommand{\Ref}[1]{\ref{#1}}

\usepackage{color, colortbl}


\title{\boldmath Factorization in deep inelastic scattering at Bjorken limit: Reduction to (1+1)D integrable models.}
\author[a,b]{H.~Babujian,}
\author[c]{M.~Karowski,}
\author[a]{and A.~Sedrakyan}
\affiliation[a]{Alikhanyan National Science Laboratory, Br.\ Alikhanian 2, Yerevan 0036, Armenia}
\affiliation[b]{Beijing Institute of Mathematical Sciences and Applications, Huairou 101408, Beijing, China}
\affiliation[c]{Freie Universit\"at Berlin, Berlin, Germany}
\abstract{We propose a Sine-Gordon description of the longitudinal sector probed in small-$x$ deep inelastic scattering on an effective string world sheet. Integrability gives nonperturbative access to the leading small-$x$ behavior through exact form factors. We test the proposal against combined HERA data with their published correlated uncertainties. In the window $Q^2>1\,\mathrm{GeV}^2$, $x\le0.01$, $y<0.3$, local powers describe the final reduced cross sections with $\chi^2/\mathrm{dof}=71.811/77$ across 19 virtuality slices. Matching the measured exponent $\gamma(Q^2)$ to the exact Sine-Gordon power defines the scale-dependent effective coupling through $\beta^2(Q^2)/\pi=16/[3-\gamma(Q^2)]$. A bounded crossover with illustrative limit $C=0.431$ is compatible with the data; the compact HERA-I $F_2$ surface fit gives $123.608/106$. We also evaluate the fixed-coupling three-particle normalization, obtaining $g_3(\infty)=0.589320245$, verify numerical convergence independently, and derive its monotonicity and asymptotic approach. The proposed QCD correspondence is distinct from these exact-model and truncated-spectral results: identifying the collective world-sheet field, electromagnetic-current matching, and scale evolution remains an open problem.}

\begin{document}
	\maketitle

\section{Introduction}

\label{sec:introduction}

Deep inelastic scattering (DIS) probes hadronic structure through the response
to a local electromagnetic current. At small Bjorken $x$, the photon--proton
energy is large and the available rapidity interval $Y=\ln(1/x)$ is long. The
hard scale $Q^{2}$ controls the transverse resolution, while the small-$x$
dependence characterizes the growth of the high-energy response. We
investigate whether this growth admits a nonperturbative description in an
integrable theory of effective longitudinal degrees of freedom.

The measured small-$x$ exponent provides the link between data and the
candidate effective theory. At each fixed $Q^{2}$, we fit $F_{2}%
(x,Q^{2})=a(Q^{2})x^{-\gamma(Q^{2})}$ over a specified window. The
normalization absorbs the remaining hard-scale dependence, while the
dimensionless slope can be compared with exact integrable-model exponents.
This separation is a phenomenological parametrization, whose domain of
validity is tested rather than assumed.

The analysis has two steps. Ratios $F_{2}(x,Q_{1}^{2})/F_{2}(x,Q_{2}^{2})$ at
common $x$ test the extent of scale separation, and fixed-$Q^{2}$ slices
determine the local logarithmic slope
\begin{equation}
\label{gamma1}\gamma(Q^{2})\simeq-\frac{\partial\ln F_{2}(x,Q^{2})}%
{\partial\ln x}\, .
\end{equation}
The HERA data show a smooth rise of this slope with $Q^{2}$. The final HERA
I+II reduced-cross-section slopes reach $0.299\pm0.008$ at $Q^{2}%
=60\,\mathrm{GeV}^{2}$ and $0.304\pm0.015$ at $Q^{2}=90\,\mathrm{GeV}^{2}$ in
the enlarged primary window. The extrapolation remains compatible with the
reference value $C=0.431$. These numbers are the quantitative bridge between
four-dimensional DIS and exact two-dimensional dynamics.

This viewpoint connects several well-established ideas. Bjorken scaling and
the parton model explain why a hard probe can resolve pointlike degrees of
freedom inside the proton~\cite{Bjorken1969,Feynman1969}. Regge theory
explains why high-energy amplitudes naturally produce powers of the energy and
hence powers of $1/x$~\cite{Regge1959,Gribov2009}. In the small-$x$ Bjorken
region these ideas meet: at fixed $Q^{2}$, the hadronic invariant mass
satisfies
\begin{equation}
\label{W}W^{2}\simeq\frac{Q^{2}}{x},
\end{equation}
so lowering $x$ opens a long high-energy interval while the hard probe remains short-distance.

QCD is the underlying four-dimensional theory. Asymptotic freedom and DGLAP
evolution describe scaling violations at fixed $x$%
~\cite{GribovLipatov-1972,GrossWilczek1973,Politzer1973,AltarelliParisi1977,Dokshitzer1977}%
. At small $x$, the BFKL/BK/JIMWLK framework organizes high-energy
logarithms~\cite{BalitskyLipatov1978,Iancu2015}. Modern factorization
treatments, including high-energy factorization and effective-field-theory
derivations, give a precise QCD setting for this
limit~\cite{CollinsSoperSterman1989,Catani1994,Neill2023}. The role of the
present analysis is to isolate the leading longitudinal power visible in the
data and to compare its exponent with exact nonperturbative results.

The link to integrability is especially useful. In the Regge limit of QCD,
Lipatov related the multicolor high-energy problem to a noncompact
$SL(2,\mathbb{C})$ spin chain~\cite{Lipatov1993,Lipatov1994}. This relation
opened a deep connection between high-energy QCD, Bethe ansatz methods, and
integrable systems~\cite{Faddeev1995,Derkachov2001,Korchemsky1995}. Recent
perturbative work continues to sharpen this high-energy
sector~\cite{Brunello2025}. A complementary nonperturbative tool is provided
by integrable quantum field theories in $(1+1)$ dimensions. Their exact $S$
matrices and exact form factors determine correlation functions with rare
analytic control~\cite{Karowski1978,Smirnov1992,Babujian1999,Lukyanov1997}.
Balog and Weisz used this method to compute DIS-like structure functions in
asymptotically free integrable models~\cite{Balog2004,Balog2005}; later work
extended the analysis to Gross--Neveu and Sine-Gordon
theories~\cite{Babujian2025}. Reviews and broader perspectives on QCD and DIS
are given in Refs.~\cite{QCD50,Zhang2022}.

We write $x\equiv x_{B}$, $Q^{2}=-q^{2}>0$, and $Y=\ln(1/x)$. The functions
$a(Q^{2}) $ and $\gamma(Q^{2})$ denote the fitted normalization and
finite-window exponent. The parameter $C$ is the candidate limiting exponent
in a phenomenological crossover. In the Sine-Gordon theory, $M$ is the mass
gap, $\beta$ is the dimensionless coupling, $\mu$ sets the potential scale,
and $g$ is a dimensionless function of $Q^{2}/M^{2}$. Its fixed-coupling
spectral calculation is separate from the fitted HERA normalization.

The primary final-HERA sample is described by local powers with $\chi
^{2}/\mathrm{dof}=71.811/77$. A compact HERA-I $F_{2}$ interpolation
constrained to the reference limit $C=0.431$ gives $123.608/106=1.166$. The
exact Sine-Gordon power, $\gamma=3-16\pi/\beta^{2}$, then supplies the
matching condition for our proposed world-sheet description. The benchmark
corresponds to $\beta^{2}/\pi=6.228$; its compatibility with the data
motivates the model without uniquely identifying it.

By ``effective longitudinal sector'' we mean the high-energy channel whose
evolution over a large rapidity interval controls the small-$x$ growth, rather
than the photon structure function $F_{L}$. We propose to represent its
collective dynamics by a Sine-Gordon field on a string world sheet. The probe
virtuality labels the matching scale, and Eq.~\eqref{beataQ} specifies the
effective coupling there. The world-sheet field and electromagnetic-current
correspondence must ultimately be derived from QCD.

The benefit of integrability is access to the leading small-$x$ behavior
through exact form factors, without a weak-coupling expansion in the effective
theory~\cite{Smirnov1992,Balog2004,Balog2005,Babujian2025}. In QCD, large
rapidity logarithms can require resummation beyond fixed order, and
nonperturbative dynamics can affect the high-energy
response~\cite{BalitskyLipatov1978,Iancu2015,AltarelliBallForte2001}. An
integrable model offers a complementary calculation; the correspondence of
that model to QCD remains the physical hypothesis. Sine-Gordon is the
candidate studied here because its coupling-dependent power is compatible with
the measured behavior.

Our contribution combines covariance-aware tests of the HERA powers with an
explicit calculation in the candidate effective model. From the published
Sine-Gordon current form factors, we evaluate the fixed-coupling
three-particle normalization, independently check the numerical integration,
and derive its monotonicity and leading asymptotic approach. These results
supply a quantitative benchmark for future current matching. The
broader-window fits and cut variations delimit the empirical support for the
exponent correspondence.

The paper is organized as follows. Section~\ref{sec:bjorken} fixes the DIS
variables and the Bjorken-limit setting. Section~\ref{sec:regge} explains the
Regge motivation for a small-$x$ power. Section~\ref{sec:exp} gives the HERA
ratio test, the final-HERA local-power fits, the Sine-Gordon $F_{2}$
representation, and the exponent crossover. Section~\ref{sec:integrable}
compares the extracted exponent with exact structure functions in
$(1+1)$-dimensional integrable models. Section~\ref{sec:summary} summarizes
the evidence for the Sine-Gordon exponent.

\section{Bjorken limit}

\label{sec:bjorken}

Bjorken scaling was proposed as a simple organizing principle for deep
inelastic scattering~\cite{Bjorken1969}. It states that in the limit of large
momentum transfer and large hadronic energy, certain dimensionless
combinations of cross sections become functions of a dimensionless scaling
variable rather than of the two large invariants separately. Feynman's parton
model gave a clear physical picture for this observation by treating the
hadron as a collection of almost free pointlike constituents during the short
interaction time~\cite{Feynman1969}.

We consider the DIS process sketched in Figs.~\ref{fig:dis_kinematics}
and~\ref{fig:optical}: an incoming lepton with momentum $p_{1}$ scatters from
a proton with momentum $p_{2}$ into an outgoing lepton with momentum $p_{3}$
and an inclusive hadronic state $X$ with total momentum $p_{4}$. The exchanged
photon carries $q=p_{1}-p_{3}$. We use
\begin{gather*}
Q^{2}\equiv-q^{2}>0,\qquad x\equiv x_{B}=\frac{Q^{2}}{2p_{2}\cdot q},\\
y=\frac{p_{2}\cdot q}{p_{2}\cdot p_{1}}=1-\frac{E_{3}}{E_{1}}.
\end{gather*}
Here $E_{1}$ and $E_{3}$ are the incoming and outgoing lepton energies in the
proton rest frame. With this notation, the Bjorken limit corresponds to
$Q^{2}\to\infty$ and $p_{2}\cdot q\to\infty$ at fixed $x$.

The hadronic invariant mass is
\[
W^{2}=(p_{2}+q)^{2}=m_{p}^{2}+Q^{2}\left( \frac{1}{x}-1\right) .
\]
Thus, at fixed $Q^{2}$ and small $x$, one has $W^{2}\simeq Q^{2}/x$. This
elementary relation is important for the rest of the paper. It shows why the
small-$x$ region is also a high-energy region and why Regge ideas can be
compared with Bjorken-limit data.

The Bjorken limit isolates short-distance information. In modern QCD language,
the relevant observables are described by the operator product expansion
together with the renormalization-group evolution of twist-two operators. QCD
predicts calculable scaling violations through asymptotic freedom and the
DGLAP evolution
equations~\cite{GribovLipatov-1972,GrossWilczek1973,Politzer1973,AltarelliParisi1977,Dokshitzer1977}%
. These logarithmic violations are a standard and well-tested part of DIS phenomenology.

For the present work, the main point is the separation of scales. The variable
$Q^{2}$ controls the transverse resolution of the virtual photon. The variable
$x$, or equivalently the rapidity interval $Y=\ln(1/x)$, controls the
available longitudinal evolution. The high-$Q^{2}$ data reveal a leading
small-$x$ structure function with a simple power law. The limiting exponent of
this power is the observable tested below.

\section{Regge poles}

\label{sec:regge}

Regge theory gives a direct language for high-energy strong
scattering~\cite{Regge1959,Gribov2009}. Instead of expanding an amplitude in
ordinary partial waves at fixed angular momentum, one analytically continues
angular momentum to a complex variable. The singularities in that complex
plane are Regge poles and Regge cuts. A Regge pole represents a whole
trajectory of hadronic states with the same quantum numbers, and its intercept
controls the leading power of the energy.

This framework is useful for DIS because small $x$ is a high-energy limit. At
fixed $Q^{2}$, the photon--proton energy satisfies $W^{2}\simeq Q^{2}/x$. Thus
decreasing $x$ increases the rapidity interval available for the hadronic
final state. A leading Regge singularity naturally turns that long rapidity
interval into a power of $1/x$. This is the physical reason why the exponent
$\gamma$ measured below is the key object: it is the effective high-energy
intercept seen by the virtual photon.

In the pole approximation the scattering amplitude can be written as
\begin{equation}
\label{regge1}A(s,t)=\sum_{i} \beta_{i}(t)\,s^{\alpha_{i}(t)},
\end{equation}
where $s$ is the center-of-mass energy squared, $t$ is the invariant momentum
transfer, $\alpha_{i}(t)$ are Regge trajectories, and $\beta_{i}(t)$ are
residues. The residue is the coupling of the external particles to the
exchanged trajectory. The leading vacuum trajectory is the Pomeron. In
perturbative QCD its small-$x$ counterpart is the BFKL
Pomeron~\cite{BalitskyLipatov1978}.

For inclusive DIS the relevant object is the forward virtual Compton
amplitude. The forward momentum transfer is $t=0$, while $Q^{2}$ is the
virtuality of the external electromagnetic currents. The hard scale $Q^{2}$
therefore enters the photon impact factors, residues, and coefficient
functions, while the Regge singularity controls the energy power. In a
one-pole picture this gives the schematic behavior
\begin{equation}
\label{regge2}F_{2}(x,Q^{2})\simeq c(Q^{2})x^{1-\alpha_{P}(0)}, \qquad
\alpha_{P}(0)=1+\epsilon,
\end{equation}
so that the structure function rises as $x^{-\epsilon}$. Regge fits to HERA
data use the same energy-power logic~\cite{Donnachie1998}. The fitted slope
$\gamma(Q^{2})$ in Sec.~\ref{sec:exp} is the measured finite-window version of
this high-energy exponent.

Regge factorization also explains why a product form is a natural first
organizing principle. The longitudinal part is governed by powers of the large
energy, or equivalently by the rapidity interval $Y=\ln(1/x)$. The transverse
part is contained in the residues and depends on the hard scale $Q^{2}$. This
is the same separation that motivates the leading form $F_{2}(x,Q^{2}%
)\approx\mathcal{F}_{2}(x)a(Q^{2})$. Multiparticle Reggeon exchange and ladder
diagrams refine this picture and generate Regge cuts, as required by
unitarity~\cite{Amati1962,Matinyan1976a,Matinyan1976b,Baier1981}.

\begin{figure}[t]
\includegraphics[width=0.7\columnwidth]{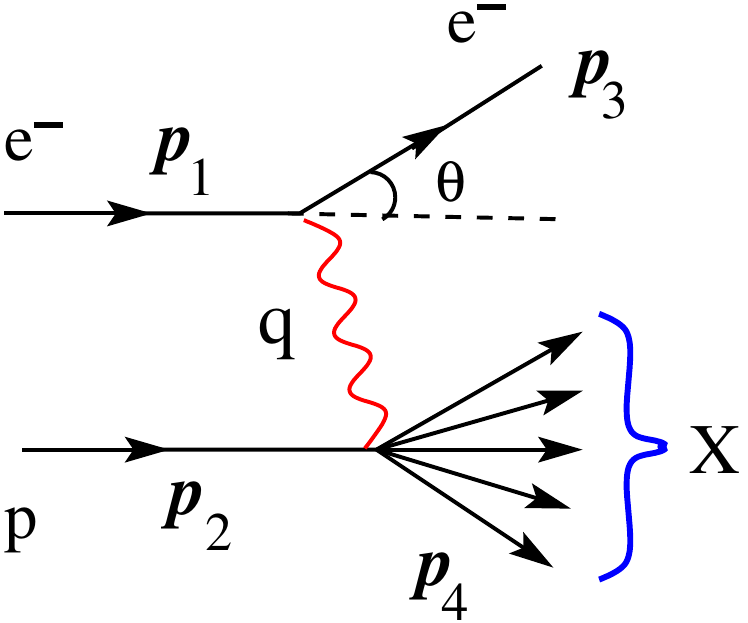}\caption{Schematic
DIS kinematics. The solid external lines denote the incoming and outgoing
lepton and the incoming proton, while the wavy line denotes the exchanged
virtual photon with momentum $q=p_{1}-p_{3}$.}%
\label{fig:dis_kinematics}%
\end{figure}

\begin{figure}[h]
\includegraphics[width=0.75\columnwidth]{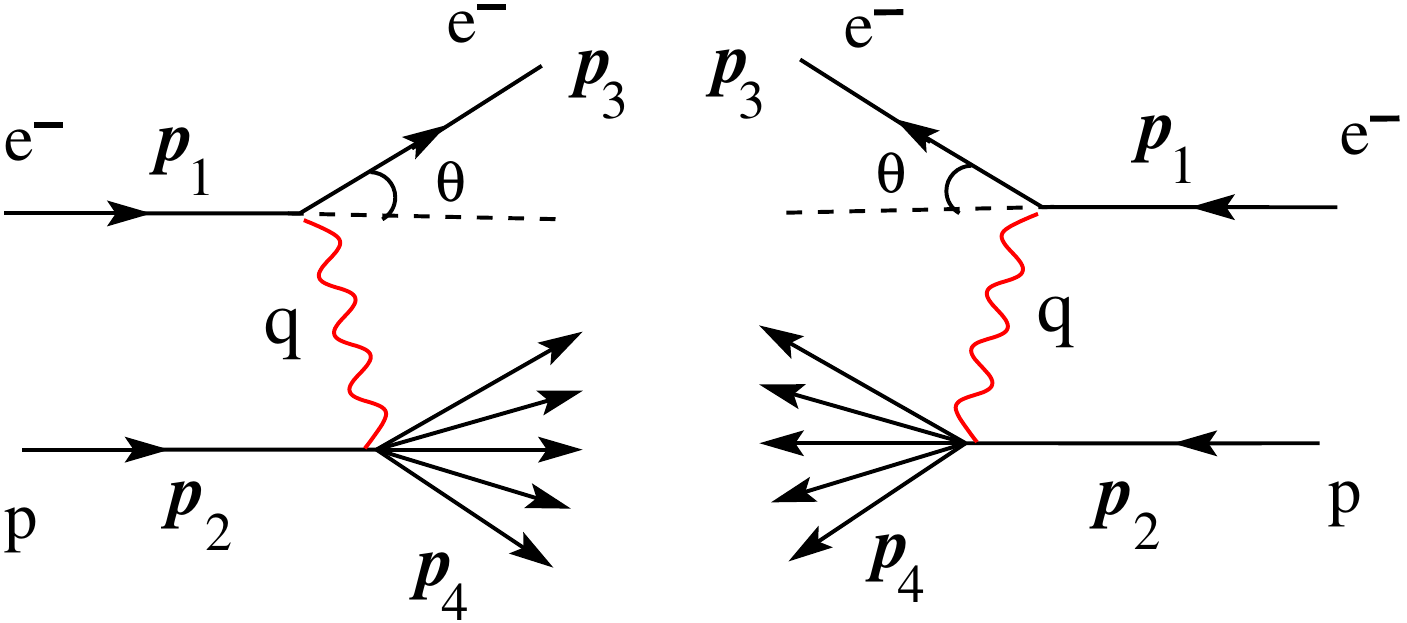}\caption{Schematic
optical-theorem representation of the DIS process in
Fig.~\ref{fig:dis_kinematics}. The solid lines denote the external states on
the two sides of the cut, and the diagram indicates how the inclusive
structure function relates to the forward amplitude.}%
\label{fig:optical}%
\end{figure}

\begin{figure}[h]
\includegraphics[width=0.72\columnwidth]{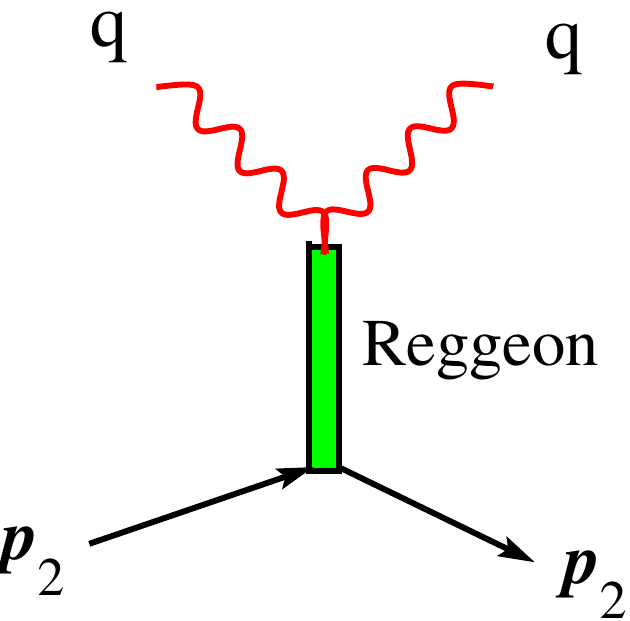}\caption{Schematic
one-Reggeon exchange contribution in the forward limit. The solid lines denote
the external states, the vertical exchanged line denotes the Reggeon, and the
wavy lines denote the virtual-photon insertions. The figure illustrates the
Regge-factorized form.}%
\label{fig:reggeon}%
\end{figure}

\begin{figure}[t]
\includegraphics[width=0.75\columnwidth]{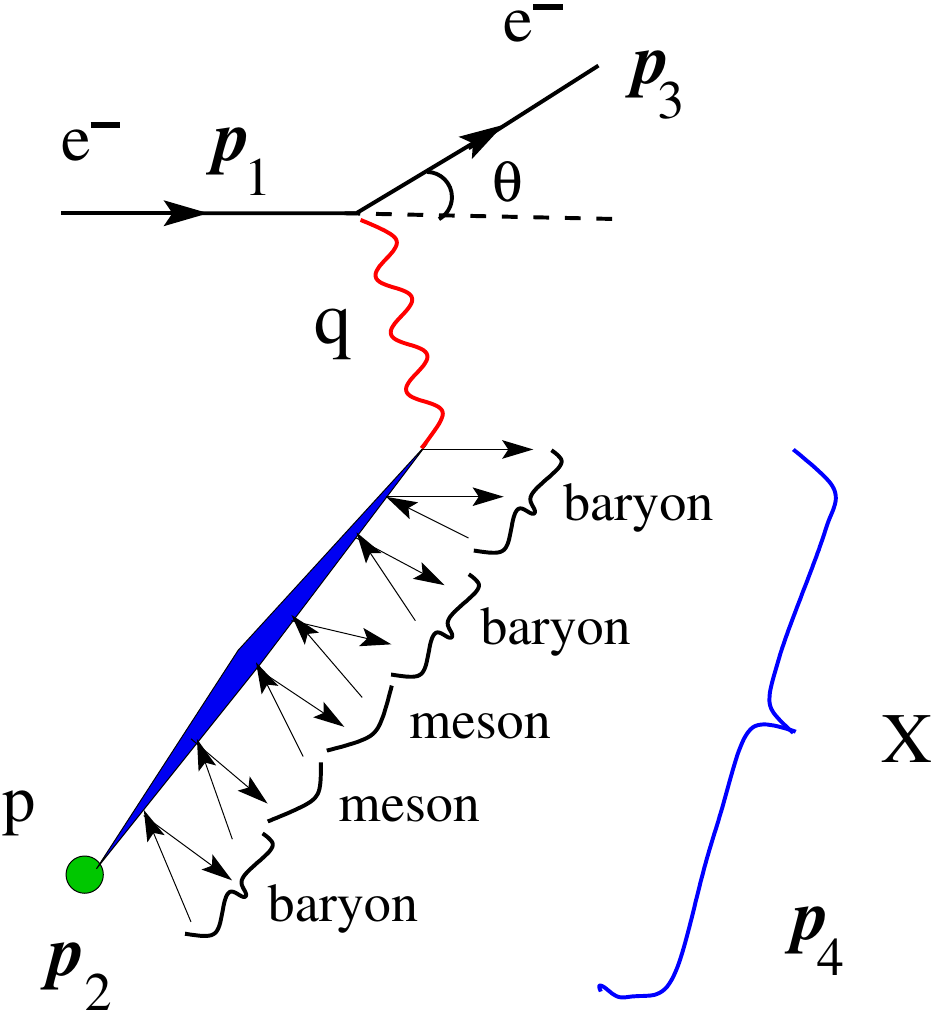}\caption{Schematic
string-inspired picture of hadron production. The solid lines indicate parton
trajectories and produced hadrons after string breaking. The figure
illustrates the separation between the hard photon vertex and the longer
longitudinal development of the final state.}%
\label{fig:string}%
\end{figure}

The diagrams in Figs.~\ref{fig:dis_kinematics}--\ref{fig:string} give the
physical content of this separation. Figure~\ref{fig:dis_kinematics} fixes the
DIS kinematics. Figure~\ref{fig:optical} shows the optical-theorem relation
between the inclusive structure function and the forward amplitude.
Figure~\ref{fig:reggeon} displays the one-Reggeon exchange channel that
produces a power of $1/x$. Figure~\ref{fig:string} gives a string-inspired
picture of the final state: the hard photon prepares the process, and the long
high-energy evolution creates the hadrons.

Modern QCD supplies precise versions of this intuition. High-energy
factorization for small-$x$ DIS was developed beyond leading order by Catani
and Hautmann~\cite{Catani1994}. Effective-field-theory methods give a clean
derivation of small-$x$ factorization in DIS~\cite{Neill2023}. Related
developments include semi-inclusive low-$x$ factorization~\cite{Boussarie2022}%
, EFT factorization for diffractive $ep$ scattering~\cite{Lee2025},
DGLAP-resummed JIMWLK evolution~\cite{Armesto2025}, and small-$x$ evolution of
gluon transverse-momentum distributions~\cite{Caucal2025}. These results show
that the Regge-to-Bjorken transition is an active and precise part of QCD.

Regge theory therefore supplies the conceptual bridge used in this paper. It
explains why the small-$x$ rise should be tested by an exponent, why the
exponent is a longitudinal observable, and why a two-dimensional effective
description can be meaningful once the hard transverse scale has been factored
into $a(Q^{2})$.

\section{Experiments and factorization}

\label{sec:exp}

Reduced unpolarized $e^{\pm}p$ cross sections are linear combinations of the
DIS structure functions; see Ref.~\cite{PDG2022} for a compact review. In the
notation fixed above, the neutral-current cross section is
\[
\label{CS1}\frac{d^{2}\sigma^{e^{\pm}p}}{dx\,dQ^{2}} =\frac{4\pi\alpha^{2}%
}{Q^{4}}\left[ (1-y)\frac{F_{2}(x,Q^{2})}{x}+y^{2}F_{1}(x,Q^{2})\right] ,
\]
where $\alpha$ is the electromagnetic fine-structure constant. At leading
twist the Callan--Gross relation is
\begin{equation}
\label{SC2}F_{2}(x,Q^{2})=2xF_{1}(x,Q^{2}).
\end{equation}
Our phenomenological discussion therefore focuses on the published
$F_{2}(x,Q^{2})$ values. In the small-$x$ high-$Q^{2}$ windows used below,
longitudinal-structure-function and electroweak corrections are already part
of the experimental extraction of $F_{2}$ in
Refs.~\cite{Aaron2009,Aaron2010,Abramowicz2015}.

\subsection{Data set and ratio test}

The analysis uses published inclusive DIS measurements from H1 and from the
combined H1--ZEUS data
sets~\cite{Adloff-2001,Aaron2009,Aaron2010,Abramowicz2015}. H1 and ZEUS were
the two general-purpose collider experiments at HERA, the electron--proton
collider at DESY. These data are ideal for the present purpose because they
cover many fixed-$Q^{2}$ slices over a wide range of small and intermediate
$x$. We use the explicit combined $F_{2}(x,Q^{2})$ table together with the
final HERA I+II reduced-cross-section table. This gives two complementary
determinations of the same local small-$x$ exponent and keeps the comparison
with the Sine-Gordon exponent direct.

The HERA measurements have small point-to-point errors in the region used
here, and many correlated uncertainties act mainly as common normalizations
inside a fixed-$Q^{2}$ slice. Such shifts change $a(Q^{2})$ more directly than
the logarithmic slope and also cancel strongly in the ratio test. This is a
useful point because the exponent is a slope in the plane $(\ln(1/x),\ln
F_{2}) $, rather than an absolute normalization. For two representative points
$i$ and $j$ in one slice, the point-to-point contribution to the slope
uncertainty is of order
\[
\delta\gamma_{ij}\simeq\frac{\left[ (\delta F_{i}/F_{i})^{2}+(\delta
F_{j}/F_{j})^{2}\right] ^{1/2}} {|\ln(x_{i}/x_{j})|}.
\]
Thus a few-percent point-to-point error spread over the logarithmic $x$
interval of the fitted slices produces a slope uncertainty of order $10^{-2}$.
The observed change of $\gamma(Q^{2})$ from about $0.15$ to about $0.30$, and
the plateau range quoted below for $C$, are therefore controlled by the data
trend rather than by an overall normalization choice. For this reason the
central-value extraction is the cleanest way to display the exponent pattern,
while the quoted stability range of $C$ records the size of the relevant slope
variation for the present comparison.

\begin{figure}[t]
\includegraphics[width=\columnwidth]{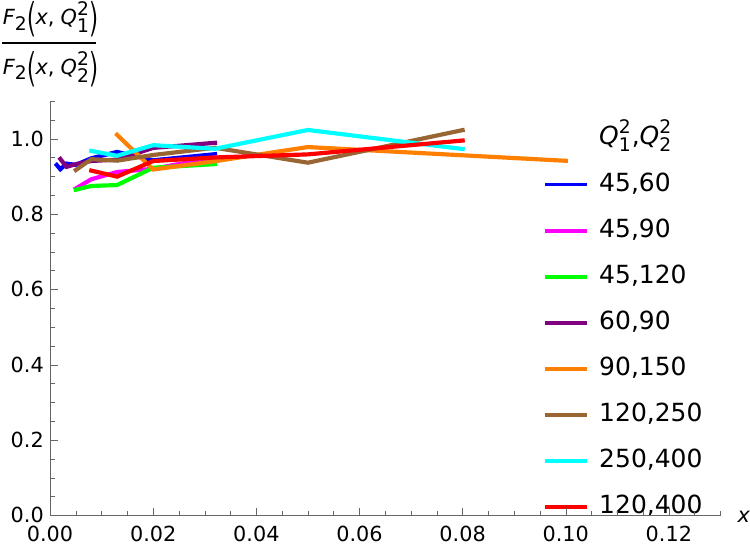}\caption{Ratios
$F_{2}(x,Q_{1}^{2})/F_{2}(x,Q_{2}^{2})$ built from the published H1 and
combined H1--ZEUS inclusive DIS points at the common $x$ values available for
the six $Q^{2}$ pairs listed in the legend. The solid colored lines connect
neighboring ratio values to display the trend. The residual drift is
compatible with scale-dependent local powers; exact product separability is
rejected over the full selected window.}%
\label{fig:ratio}%
\end{figure}

For the ratio test in Fig.~\ref{fig:ratio} we use the eight $Q^{2}$ pairs
shown in the legend: $(45,60)$, $(45,90)$, $(45,120)$, $(60,90)$, $(90,150)$,
$(120,250)$, $(250,400)$, and $(120,400)\,\mathrm{GeV}^{2}$. For each pair we
use common $x$ values available in the published data. For the exponent fits
we use the same small-$x$ local-power method in the explicit $F_{2}$ table and
in the final HERA I+II reduced-cross-section table. The exponent is always the
finite-window local slope of the rising part of a fixed-$Q^{2}$ slice. This is
the finite-energy way in which HERA approaches the large-$Q^{2}$ small-$x$ slope.

If the structure function factorizes as
\begin{equation}
\label{SF-1}F_{2}(x,Q^{2})=\mathcal{F}_{2}(x)\,a(Q^{2}),
\end{equation}
then the ratio at two hard scales is
\[
\frac{F_{2}(x,Q^{2})}{F_{2}(x,Q^{\prime2})}=\frac{a(Q^{2})}{a(Q^{\prime2})}%
\]
and is therefore independent of $x$. With the local-power form used below one
obtains the more detailed relation
\begin{equation}
\label{fac-1}\frac{F_{2}(x,Q_{1}^{2})}{F_{2}(x,Q_{2}^{2})} =\frac{a(Q_{1}%
^{2})}{a(Q_{2}^{2})}\, x^{-[\gamma(Q_{1}^{2})-\gamma(Q_{2}^{2})]}.
\end{equation}
Thus exact factorization is recovered when the slopes coincide, while the
observed slow change of $\gamma(Q^{2})$ gives only a mild residual $x$
dependence over the selected windows. For example, a difference $\Delta
\gamma=0.03$ changes a ratio by only $10^{0.03}\simeq1.07$ across one decade
in $x$. This is the level of residual drift visible in Fig.~\ref{fig:ratio}.
The ratio test and the fixed-$Q^{2}$ slope extraction are therefore two
consistent views of the same leading pattern.

Figure~\ref{fig:ratio} gives the first verification of the central idea. Each
ratio curve changes slowly with $x$ over the displayed interval, even when the
two $Q^{2}$ values are well separated. The observed stability is the data
basis for fitting each fixed-$Q^{2}$ slice by a local power. In this paper
``factorization'' denotes this leading asymptotic separation of the
longitudinal small-$x$ dependence from the hard-scale normalization, with the
remaining finite-$Q^{2}$ change encoded in the slowly varying local exponent.
The ratio test is important because it tests the premise before the exponent
is extracted: the exponent comparison with Sine-Gordon theory is meaningful
only when the leading $x$ dependence can be isolated from the overall $Q^{2}$ normalization.

\subsection{Local-power exponent extraction}

Motivated by the ratio test, we fit the low-$x$ part of each fixed-$Q^{2}$
slice by
\begin{equation}
F_{2}(x,Q^{2})=\frac{a(Q^{2})}{x^{\gamma(Q^{2})}}.\label{eq:ansatz}%
\end{equation}
Here $\gamma(Q^{2})$ is the finite-window local logarithmic slope,
\[
\gamma(Q^{2})\simeq-\frac{\Delta\ln F_{2}(x,Q^{2})}{\Delta\ln x},
\]
extracted from the power-law part of the fixed-$Q^{2}$ slice. It is the
measured exponent that controls the small-$x$ rise in that slice.

The same definition is used for the explicit HERA $F_{2}$ table and for the
final HERA I+II reduced-cross-section proxy. The representative final-HERA
slices shown below display the straight log-log behavior from which the local
exponent is obtained. The full set of extracted exponents is summarized in
Fig.~\ref{fig:gamma}, which is the main exponent plot used in the Sine-Gordon
comparison. The exponent grows from the soft finite-$Q^{2}$ value toward the
high-$Q^{2}$ plateau near $0.43$. Physically, $\gamma(Q^{2})$ measures the
rate at which the inclusive longitudinal response grows when the rapidity
interval $Y=\ln(1/x)$ is increased: an increment $Y\to Y+\Delta Y$ multiplies
the leading contribution by approximately $\exp[\gamma(Q^{2})\Delta Y]$.

\subsection{Final-HERA numerical confirmation}

We next add a cross-check with the overlapping final-HERA combination using
the official public H1--ZEUS combined data
tables~\cite{H1ZEUSCombinedResults,HERAPDF20web}. This check uses the final
HERA I+II neutral-current $e^{+}p$ reduced cross sections at $E_{p}%
=920\,\mathrm{GeV}$, in addition to the earlier explicit $F_{2}$ extraction.
The purpose is simple: we quantify the finite-window factorized form and
extract the local small-$x$ exponent with a larger and more precise HERA data set.

For this additional test we selected points with $Q^{2}>1\,\mathrm{GeV}^{2}$,
$y<0.3$, and $x\le0.01$, with at least three distinct $x$ points per
fixed-$Q^{2}$ slice. The cut $y<0.3$ reduces the sensitivity of the reduced
cross section $\sigma_{r}$ to the longitudinal-structure-function correction,
so the logarithmic slope of $\sigma_{r}$ is a stable proxy for the small-$x$
slope of $F_{2}$. The selected samples contain 114 HERA-I $F_{2}$ points and
115 final-HERA reduced-cross-section points, separately fitted in 19 slices
over $1.2\le Q^{2}\le90\,\mathrm{GeV}^{2}$. The samples overlap and are not
combined into one likelihood. Then each fixed-$Q^{2}$ slice was fitted by the
same local power law used above,
\begin{equation}
D(x,Q^{2})=a(Q^{2})x^{-\gamma(Q^{2})},\label{eq:blue_local_power}%
\end{equation}
where $D$ denotes either $F_{2}$ or $\sigma_{r}$ in the corresponding table.
The fit was performed in logarithmic variables with the full fixed covariance
constructed from published statistical, uncorrelated and signed correlated
errors. The following diagonal-error notation is retained; numerically its
generalized least-squares extension $\chi^{2}=\boldsymbol{r}^{T}%
V^{-1}\boldsymbol{r}$ is used, without adding the quoted total uncertainty a
second time:
\begin{equation}%
\begin{split}
\chi^{2}(Q^{2}) & =\sum_{i} \left[  \frac{\ln D_{i}-\ln a(Q^{2})+\gamma
(Q^{2})\ln x_{i}}{\delta_{i}} \right] ^{2},\\
\delta_{i} & =\frac{\Delta_{i}}{100}.
\end{split}
\label{eq:blue_log_chi2}%
\end{equation}
where $\Delta_{i}$ is the quoted total relative uncertainty of the point. This
direct slope test is designed for the exponent. Correlated normalization
errors mainly change $a(Q^{2})$, while the exponent $\gamma(Q^{2})$ is
controlled by point-to-point changes with $x$.

For the global model comparison we quote the reduced statistic
\begin{equation}
\chi^{2}_{\nu}=\frac{\chi^{2}}{\nu}=\frac{\sum_{i} r_{i}^{2}}{N-p}%
,\label{eq:green_reduced_chi}%
\end{equation}
where $r_{i}$ is a covariance-whitened log-space residual, $N$ is the number
of fitted points, and $p$ is the rank of the corresponding linear fit in
logarithmic variables. For the local-power model this is the same as summing
the fixed-$Q^{2}$ slice chi-squares and dividing by the summed number of slice
degrees of freedom. Values close to unity show that the local-power
description captures the measured slope information with the published
correlated uncertainties propagated to logarithmic space.




\begin{figure}[t]
\centering
\includegraphics[width=\columnwidth]{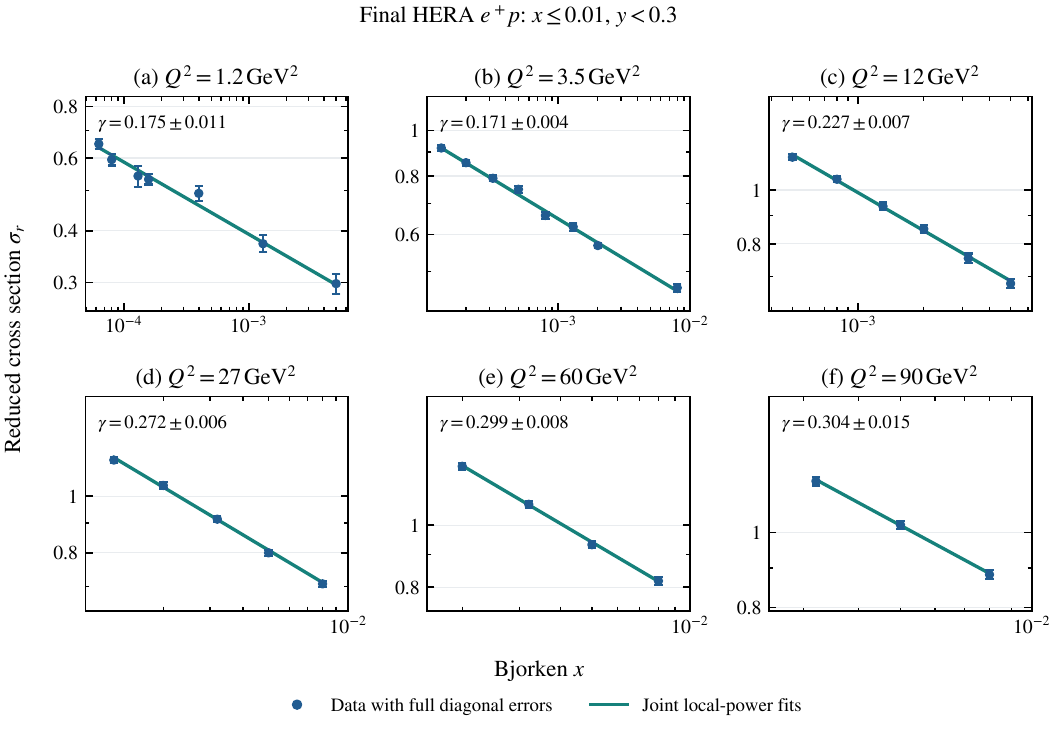}\caption{Representative
final-HERA fixed-$Q^{2}$ slices and joint local-power fits for $Q^{2}%
>1\,\mathrm{GeV}^{2}$, $x\le0.01$, $y<0.3$. Bars show full diagonal errors;
fits retain the correlations.}%
\label{fig:blue_hera_slices}%
\end{figure}



The numerical outcome is shown in Fig.
~\ref{fig:blue_hera_slices}. The hierarchy is clear: the finite-$Q^{2}$
local-power form gives global reduced values $\chi^{2}_{\nu}=71.811/77=0.933$
for the final HERA reduced cross sections and $\chi^{2}_{\nu}=71.043/76=0.935$
for HERA-I $F_{2}$. Thus the high-$Q^{2}$, small-$x$ rise is accurately
described by a local power whose exponent changes slowly with $Q^{2}$.

The enlarged-$Q^{2}$-range check also supports the local exponent used in the
Sine-Gordon comparison. From the final HERA I+II reduced-cross-section table
we find $\gamma(1.5\,\mathrm{GeV}^{2})=0.151\pm0.009$, $\gamma
(12\,\mathrm{GeV}^{2})=0.227\pm0.007$, $\gamma(60\,\mathrm{GeV}^{2}%
)=0.299\pm0.008$, and $\gamma(90\,\mathrm{GeV}^{2})=0.304\pm0.015$. The
primary window ends at $90\,\mathrm{GeV}^{2}$; its limiting value is therefore
a crossover extrapolation.

\subsection{Sine-Gordon finite-HERA representation of the structure function}

\label{subsec:best_f2}

The exponent extracted above suggests a direct physics representation of the
full small-$x$ surface. We use
\begin{equation}
F_{2}^{\mathrm{SG}}(x,Q^{2})=\exp[P_{4}(T)]\,x^{-\gamma_{\mathrm{SG}}(Q^{2}%
)},\label{eq:best_sg_F2}%
\end{equation}
where
\begin{equation}
T=\ln\!\left( \frac{Q^{2}}{100\,\mathrm{GeV}^{2}}\right) ,
\end{equation}
and
\begin{equation}
\gamma(Q^{2})=\gamma_{\mathrm{soft}}+ (C-\gamma_{\mathrm{soft}}) \frac
{(Q^{2}/Q_{*}^{2})^{\omega}}{1+(Q^{2}/Q_{*}^{2})^{\omega}}%
.\label{eq:sg_crossover_F2}%
\end{equation}
For the benchmark fit within the proposed description, we fix the illustrative
limiting exponent to
\begin{equation}
C=\gamma_{\mathrm{SG}}(\infty)=0.431 .
\end{equation}
The explicit HERA $F_{2}$ fit gives
\begin{equation}
\gamma_{\mathrm{soft}}=0.105389, \qquad Q_{*}^{2}=26.86153\,\mathrm{GeV}^{2},
\qquad\omega=0.610136,\label{eq:sg_crossover_numbers_f2}%
\end{equation}
and the smooth normalization is
\begin{align}
P_{4}(T)= & -1.690208-0.003182T+0.068122T^{2}\nonumber\\
& +0.026721T^{3}+0.001995T^{4}.\label{eq:sg_norm_p4}%
\end{align}
This Sine-Gordon-constrained expression gives
\begin{equation}
\chi^{2}_{\nu}=\frac{123.608}{106}=1.166 .\label{eq:sg_full_surface_chi}%
\end{equation}
The value is close to unity over the selected HERA window. This is the central
finite-$Q^{2}$ result: the selected $F_{2}$ surface is described by a local
power with a crossover constrained to the illustrative Sine-Gordon exponent,
while its normalization is interpolated by $P_{4}(T)$. The limiting value is
an imposed benchmark compatible with the data, not a unique consequence of
Sine-Gordon theory or a direct measurement of an asymptotic plateau.

The fit has a simple physical interpretation within the proposal. The exponent
is the longitudinal matching observable. The fitted normalization absorbs
hard-scale dependence phenomenologically, including effects that a derivation
would allocate to the four-dimensional current and transverse dynamics. The
parameters $\gamma_{\mathrm{soft}}$, $Q_{*}^{2}$, and $\omega$ describe the
finite-HERA crossover. Mapping this exponent through Eq.~\eqref{beataQ}
specifies a scale-dependent effective world-sheet coupling; it does not by
itself establish a universal ultraviolet limit.

\begin{figure}[t]
\centering
\includegraphics[width=\columnwidth]{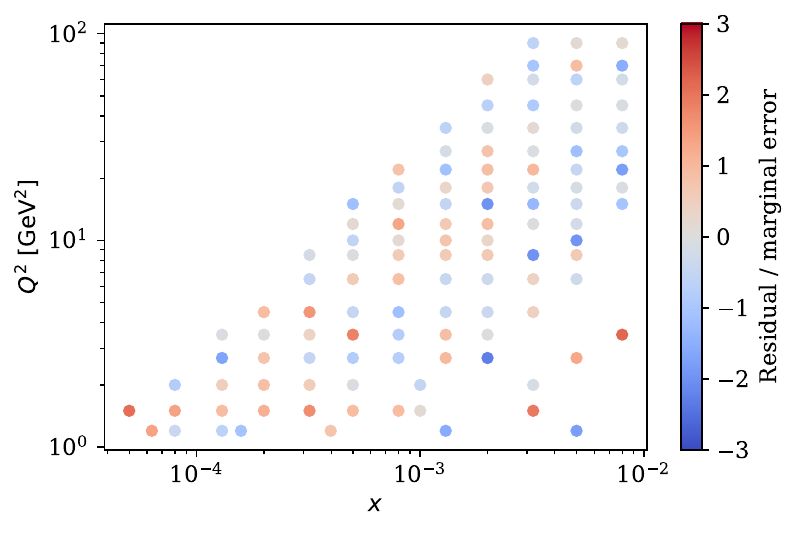}\caption{Residuals
of the compact $P_{4}$ surface fit to explicit HERA-I $F_{2}$ in
Eq.~\eqref{eq:best_sg_F2}, with $Q^{2}>1\,\mathrm{GeV}^{2}$, $x\le0.01$,
$y<0.3$. The fit gives $123.608/106=1.166$. Plotted residuals are divided by
marginal errors; the fit statistic uses the full covariance.}%
\label{fig:f2_bestfit_scan}%
\end{figure}

\begin{figure}[t]
\centering
\includegraphics[width=\columnwidth]{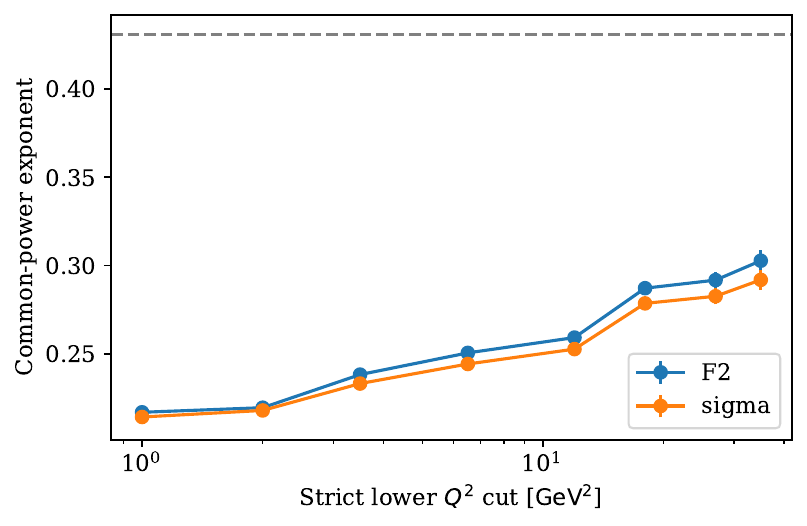}\caption{Common-power
exponent after raising the strict lower $Q^{2}$ cut within $x\le0.01$,
$y<0.3$. The reference line is $C=0.431$; the controlled sample ends at
$90\,\mathrm{GeV}^{2}$.}%
\label{fig:highQ2_universality_scan}%
\end{figure}

\subsection{Exponent verification}

\begin{figure}[t]
\includegraphics[width=\columnwidth]{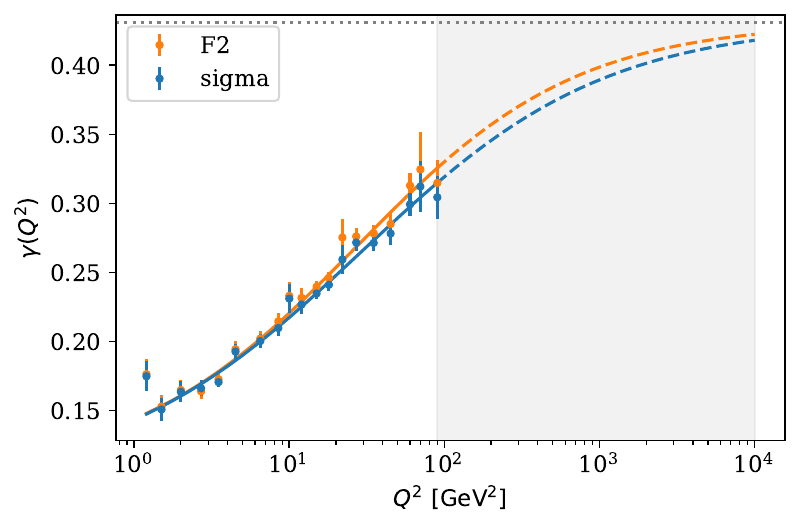}\caption{Local
HERA-I $F_{2}$ and final-HERA $\sigma_{r}$ exponents for the strict
$Q^{2}>1\,\mathrm{GeV}^{2}$ sample. Solid curves are the compact $P_{4}$
surface crossovers with $C=0.431$. Dashed extensions above $90\,\mathrm{GeV}%
^{2}$ are extrapolations. Bars are marginal slope errors.}%
\label{fig:gamma}%
\end{figure}

The values of $\gamma(Q^{2})$ extracted from fixed-$Q^{2}$ slices are
summarized in Fig.~\ref{fig:gamma}. We use a bounded crossover as a
phenomenological interpolation compatible with a candidate limiting exponent.
The approximately linear dependence on $\ln Q^{2}$ over the measured interval
can be represented by the middle part of that crossover. The finite data do
not establish its asymptotic endpoint.

A QCD renormalization-group argument motivates treating the exponent as scale
dependent, although it does not derive the specific bounded crossover used
here. At small $x$ we write the leading contribution as
\begin{equation}
F_{2}(x,Q^{2})\simeq A(Q^{2})\exp[\gamma(Q^{2})Y], \qquad Y=\ln\frac{1}{x}
.\label{eq:qcd_local_intercept}%
\end{equation}
Taking a derivative with respect to the hard-scale variable $t=\ln Q^{2}$
gives
\begin{equation}
\frac{\partial\ln F_{2}}{\partial t} =\frac{d\ln A}{dt}+Y\frac{d\gamma}{dt}
.\label{eq:qcd_slope_rg_identity}%
\end{equation}
The same statement can be written in Mellin space. Define
\[
F_{2}(N,Q^{2})=\int_{0}^{1} dx\,x^{N-1}F_{2}(x,Q^{2}).
\]
For an ideal asymptotic power $F_{2}\sim x^{-\gamma(Q^{2})}$, the Mellin
transform has a singularity at $N=\gamma(Q^{2})$. A finite-window fit does not
alone establish that analytic structure. In perturbative small-$x$ evolution
this singularity is fixed schematically by a kernel condition
\[
\Phi(N,\alpha_{s}(Q^{2}),Q^{2})=0, \qquad N=\gamma(Q^{2}),
\]
where $\Phi$ represents the resummed high-energy anomalous-dimension or dipole
kernel. Differentiating this condition gives the first-order flow
\[
\frac{d\gamma}{dt}=- \frac{\partial_{t}\Phi+\beta(\alpha_{s})\partial
_{\alpha_{s}}\Phi} {\partial_{N}\Phi}\Bigg|_{N=\gamma(Q^{2})} .
\]
This relation motivates a scale-dependent exponent but leaves its evolution
dependent on the explicit QCD kernel. We therefore use the bounded crossover
as an interpolation, rather than as a derived QCD flow.

Within the proposed description, Eq.~\eqref{beataQ} encodes the exponent
evolution in a matched world-sheet coupling. We choose a crossover with
endpoints $\gamma_{\mathrm{soft}}$ and $\gamma_{\mathrm{SG}}$. Linearization
of the assumed flow near those endpoints gives
\begin{equation}
\begin{aligned} \frac{d}{dt}(\gamma-\gamma_{\rm soft}) &=\omega(\gamma-\gamma_{\rm soft}),\\ \frac{d}{dt}(\gamma_{\rm SG}-\gamma) &=-\omega(\gamma_{\rm SG}-\gamma). \end{aligned}\label{eq:qcd_fixed_point_linearization}%
\end{equation}
to leading order in the distance from each endpoint. A minimal one-variable
interpolation with these endpoints and the required monotonicity is given
below; its identification with a world-sheet beta function would require
further matching.

Let $t=\ln(Q^{2}/Q_{*}^{2})$ and define the normalized crossover coordinate
\begin{equation}
\eta(t)=\frac{\gamma(Q^{2})-\gamma_{\mathrm{soft}}} {\gamma_{\mathrm{SG}%
}-\gamma_{\mathrm{soft}}} .\label{eq:green_eta_def}%
\end{equation}
We use the logistic crossover equation
\begin{equation}
\frac{d\eta}{dt}=\omega\eta(1-\eta),\label{eq:green_logistic_rg}%
\end{equation}
whose solution is
\begin{equation}
\gamma(Q^{2})=\gamma_{\mathrm{soft}}+ (\gamma_{\mathrm{SG}}-\gamma
_{\mathrm{soft}}) \frac{(Q^{2}/Q_{*}^{2})^{\omega}}{1+(Q^{2}/Q_{*}%
^{2})^{\omega}}\label{eq:gamma_fit}%
\end{equation}
This is the finite-$Q^{2}$ scaling function used below. The parameter
$\gamma_{\mathrm{SG}}=\gamma(\infty)=C$ is the asymptotic exponent to be
matched to Sine-Gordon theory. The quantities $\gamma_{\mathrm{soft}}$,
$Q_{*}^{2}$, and $\omega$ describe the HERA crossover into the asymptotic
regime. Near $Q^{2}=Q_{*}^{2}$, Eq.~\eqref{eq:gamma_fit} gives
\begin{equation}
\gamma(Q^{2})\simeq\frac{\gamma_{\mathrm{soft}}+\gamma_{\mathrm{SG}}}{2}
+\frac{\omega(\gamma_{\mathrm{SG}}-\gamma_{\mathrm{soft}})}{4} \ln\frac{Q^{2}%
}{Q_{*}^{2}},\label{eq:green_linearized}%
\end{equation}
which explains why Fig.~\ref{fig:gamma} appears nearly straight over a finite
logarithmic interval. At large $Q^{2}$ the same formula gives
\begin{equation}
\gamma_{\mathrm{SG}}-\gamma(Q^{2})\simeq(\gamma_{\mathrm{SG}}-\gamma
_{\mathrm{soft}}) \left( \frac{Q_{*}^{2}}{Q^{2}}\right) ^{\omega
},\label{eq:green_large_q2}%
\end{equation}
so the approach to the limiting Sine-Gordon exponent is still power-like
asymptotically, while preserving the observed finite-HERA crossover over the
measured range.

The same bounded crossover has now been fitted directly to the two HERA
surfaces used in this paper. For the explicit HERA $F_{2}$ table, with the
limiting value fixed to $C=0.431$, the fit gives
\begin{equation}
\begin{aligned} \gamma_{\rm soft}&=0.105389,\qquad Q_*^2=26.86153\,\mathrm{GeV}^2,\\ \omega&=0.610136,\qquad \chi^2_\nu=\frac{123.608}{106}=1.166 . \end{aligned}\label{eq:green_2010_surface_fit}%
\end{equation}
For the final HERA I+II $\sigma_{r}$ proxy surface, the same Sine-Gordon limit
gives
\begin{equation}
\begin{aligned} \gamma_{\rm soft}&=0.098036,\qquad Q_*^2=28.86314\,\mathrm{GeV}^2,\\ \omega&=0.549761,\qquad \chi^2_\nu=\frac{125.370}{107}=1.172 . \end{aligned}\label{eq:green_2015_surface_fit}%
\end{equation}
Thus both data sets prefer the same physical picture: a finite-crossover scale
$Q_{*}^{2}\simeq27$--$29\,\mathrm{GeV}^{2}$ and an ultraviolet limiting
exponent fixed by the Sine-Gordon value. Around the midpoint of the crossover
the local slope is approximately $\omega(\gamma_{\mathrm{SG}}-\gamma
_{\mathrm{soft}})/4\simeq0.050$, close to the direct linear diagnostic slope
in $\ln Q^{2}$. This explains why Fig.~\ref{fig:gamma} appears nearly straight
over the measured HERA interval even though the asymptotic function is bounded.

When the limiting exponent is also left free in Eq.~\eqref{eq:gamma_fit}, the
final-HERA exponent diagnostic with independent slice normalizations gives
$C=0.3305\pm0.0261$, with a broad profile reflecting the finite HERA lever arm
above the crossover scale. The Sine-Gordon value $C=0.431$ has $\Delta\chi
^{2}=2.90$ relative to that free minimum, and the conditional profile interval
$\Delta\chi^{2}\le3.84$ gives $0.298<C<0.483$. The fixed-reference fit gives
$\chi^{2}/\nu=91.748/93$ for final HERA and $91.175/92$ for HERA-I $F_{2}$.
These diagnostics use independent slice normalizations, separately from the
compact quartic-normalization surface fits quoted above. Under the Sine-Gordon
relation $C=3-16\pi/\beta^{2}$, the fixed value $C=0.431$ gives $\beta^{2}%
/\pi=6.228$, while the conditional profile interval maps to $5.922<\beta
^{2}/\pi<6.357$. This entire interval lies inside the repulsive solitonic
regime. The reference Sine-Gordon value is therefore compatible with the
finite-$Q^{2}$ consistency test.

\begin{figure}[t]
\includegraphics[width=\columnwidth]{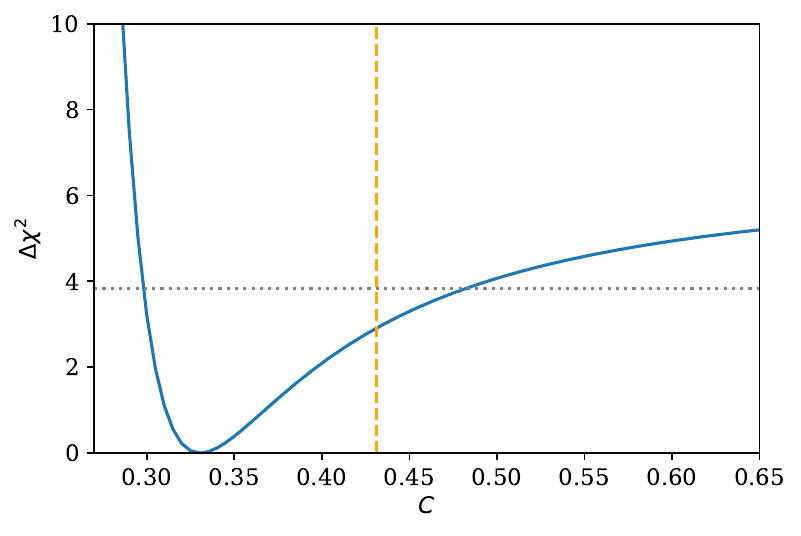}\caption{Conditional
final-HERA limiting-exponent profile for $Q^{2}>1\,\mathrm{GeV}^{2}$,
$x\le0.01$, $y<0.3$, with independent slice normalizations. The reference
$C=0.431$ costs $\Delta\chi^{2}=2.90$. The conditional $3.84$ range is
approximately $0.298<C<0.483$.}%
\label{fig:green_C_profile}%
\end{figure}

The normalization is treated through the polynomial $P_{4}(T)$ in
Eq.~\eqref{eq:sg_norm_p4}. The measured exponent fixes the proposed coupling
prescription, while the normalization interpolates the remaining hard-scale
dependence. A quantitative world-sheet interpretation of that normalization
requires the electromagnetic-current matching and effective mass scale. The
constrained expression in Eq.~\eqref{eq:best_sg_F2} fits the selected surface
with $\chi^{2}_{\nu}=1.166$; this is phenomenological support for the
proposal, rather than a derivation of the matching coefficients.

The resulting formula is
\begin{align}
& F_{2}^{\mathrm{SG}}(x,Q^{2})=\exp[P_{4}(T)]\,x^{-\gamma(Q^{2})}%
,\label{eq:asymptotic_fit}\\
& T=\ln\!\left( \frac{Q^{2}}{100\,\mathrm{GeV}^{2}}\right) , \quad
R=\frac{Q^{2}}{26.86153\,\mathrm{GeV}^{2}},\nonumber\\
& \gamma(Q^{2})=0.105389+0.325611\, \frac{R^{0.610136}}{1+R^{0.610136}%
}.\label{eq:green_gamma_cross_num}%
\end{align}
where $0.325611=0.431-0.105389$ and $P_{4}(T)$ is given in
Eq.~\eqref{eq:sg_norm_p4}. Equation~\eqref{eq:asymptotic_fit} is the final
phenomenological form used in the paper. It keeps the original small-$x$ power
logic and uses a bounded finite-$Q^{2}$ approach to the Sine-Gordon plateau.
The data-driven profile gives the plateau interval $0.298<C<0.483$, reflecting
the finite HERA lever arm above the crossover scale. The Sine-Gordon value
$C=0.431$ lies inside this conditional interval and is compatible with the
extrapolated crossover.

Recent small-$x$ QCD analyses provide useful benchmarks. Small-$x$ resummation
has long been tested against HERA structure-function
data~\cite{AltarelliBallForte2001}. Collinearly improved BK evolution, NLO
dipole calculations with massive quarks, and NLO BK initial-condition fits
have also been confronted with HERA
data~\cite{Iancu2015,Hanninen2023,Casuga2025}. These approaches give the full
QCD setting and describe the same HERA data from the four-dimensional
perturbative side. The exponent test is complementary: it isolates the leading
longitudinal power and makes it possible to compare that one dimensionless
number with exact $(1+1)$-dimensional results.

\section{Structure functions in \texorpdfstring{$(1+1)$}{(1+1)}-dimensional
integrable models}

\label{sec:integrable}

\subsection{Exact form-factor setting}

Integrable quantum field theories in $(1+1)$ dimensions provide exact
nonperturbative laboratories for structure functions. This is the point where
the measured exponent becomes a theory test. The reason is special to two
dimensions. Multiparticle scattering factorizes into two-body scattering,
elastic channels have exact asymptotic states, and the exact $S$ matrix fixes
the analytic structure of form factors. Local-operator matrix elements can
then be built explicitly and inserted into spectral
sums~\cite{Karowski1978,Smirnov1992,Babujian1999,Lukyanov1997}.

The physical parallel with small-$x$ DIS is the role of rapidity. In
four-dimensional DIS the small-$x$ limit opens a large longitudinal rapidity
interval, while $Q^{2}$ sets the hard transverse scale. In a two-dimensional
massive theory, rapidity differences organize the exact form-factor expansion.
This motivates using an integrable world-sheet model for the longitudinal
growth. The proposed correspondence is tested here through an
exponent-matching condition; an operator correspondence to the QCD
electromagnetic current remains to be derived.

For a local operator $\mathcal{O}$, the analogue of a DIS structure function
is obtained from the current commutator
\begin{equation}
W_{\mathcal{O}}(p,q)=\int d^{2}z\,e^{iq\cdot z}\langle p|[\mathcal{O}%
(z),\mathcal{O}(0)]|p\rangle.\label{eq:formfactor_object}%
\end{equation}
Here $|p\rangle$ is a one-particle asymptotic state of mass $M$, $q$ is the
momentum carried by the probe, and the two-dimensional Bjorken variable is
$x=Q^{2}/(2p\cdot q)$ with $Q^{2}=-q^{2}>0$. The coordinate is denoted by $z$
to keep it distinct from Bjorken $x$.

The form-factor expansion follows by inserting a complete set of exact
asymptotic states between the two operators in
Eq.~\eqref{eq:formfactor_object}. Each intermediate state is labeled by
rapidities. The limit $x\to0$ corresponds to large rapidity separations. The
leading power of $x$ is therefore fixed by the leading high-rapidity behavior
of the exact form factors. This is why the exponent is especially robust: once
the two-dimensional theory and the operator are chosen, the leading small-$x$
power follows from the exact analytic structure.

Once the integrable theory, operator, and coupling are specified, the exact
high-rapidity form factors determine its leading small-$x$ exponent
nonperturbatively. The normalization requires a spectral sum: the
three-particle contribution calculated below is a controlled numerical
evaluation of that term, rather than an exact evaluation of the complete response.

For the $O(3)$ sigma model, and more generally for the asymptotically free
rational-$S$-matrix models studied by Balog and Weisz, the result is
\begin{equation}
F_{2}(x,Q^{2})\xrightarrow[x\to 0]{}a(Q^{2})\,\frac{1}{x\ln^{2}(1/x)}%
.\label{eq:universal_af}%
\end{equation}
This universal law is factorized and contains a logarithmic correction to the
$1/x$ rise. The wider analysis of Ref.~\cite{Babujian2025} separates the known
examples into two broad groups. The first contains asymptotically free models
with rational $S$ matrices, including the $O(N)$ sigma model and the $SU(N)$
Gross--Neveu model~\cite{GrossNeveu1974}. The second contains models with
$U(1)$ or $Z(N)$ symmetry, including Sine-Gordon, sinh-Gordon or , and $Z(N)$
Ising models. In this second group the small-$x$ exponent depends on the
coupling or on the discrete parameter of the model.

The Sine-Gordon model alias the massive Thirring model is defined by the
Lagrangians
\begin{eqnarray}
\label{SG1}
&&\mathcal{L}^{SG}=\frac{1}{2}(\partial_{\mu}\phi)^{2}+\frac{\alpha}{\beta^{2}%
}(\cos\beta\phi-1),\nonumber\\
&&\partial^{a}\partial_{a}\varphi+\frac{\mu^{2}}{\beta}\sin(\beta\varphi)=0,\\
&&\mathcal{L}^{MTM}=\bar{\psi}(i\gamma\partial-M)\psi-\frac{1}{2}g(\bar{\psi
}\gamma^{\mu}\psi)^{2},\nonumber
\end{eqnarray}
respectively. The Fermi field $\psi$ correspond to the soliton and antisoliton
and the bose field $\phi$ to the lowest `breather' which is the lowest soliton
antisoliton bound state. The precise relation between the related coupling
constants was found by Coleman \cite{Coleman-1973} within the framework of perturbation
theory
\[
\nu=\frac{\beta^{2}}{8\pi-\beta^{2}}=\frac{\pi}{\pi+2g}%
\]
where the parameter $\nu$ is introduced for later convenience,
$\partial^{a}\partial_{a}$ is the two-dimensional Laplacian, $\beta$
is the dimensionless coupling and $\mu$ sets the mass scale.

In the repulsive regime $\beta^{2}>4\pi$, the spectrum is the
soliton--antisoliton sector. A soliton is a topological kink of the field, and
an antisoliton is the kink with opposite topological charge. In that regime
the exact form-factor analysis gives~\cite{Babujian2025}

\begin{align}
F_{2}(x,Q^{2}) &  \xrightarrow[x\to 0]{}4\sin^{2}\!\left(  \frac{\pi}{\nu
}\right)  g(Q^{2})x^{-\gamma},\label{eq:sg_power}\\
\gamma &  =1-\frac{2}{\nu}=3-\frac{16\pi}{\beta^{2}},\qquad\nu=\frac{\beta
^{2}}{8\pi-\beta^{2}}>1.\label{eq:sg_gamma}%
\end{align}
The function $g(Q^{2})$ is a dimensionless function of $Q^{2}/M^{2}$ defined
by a sum over multiparticle form-factor contributions. The exponent is fixed
by the leading high-rapidity asymptotics. The normalization records the
strength of the current matrix elements and is shown below through the
three-particle intermediate-state approximation. The square on the sine is
important only for the normalization; the comparison with HERA is controlled
by the exponent.

\subsection{Comparison with the HERA fit}

The comparison is direct and highly discriminating. The exact small-$x$
functions fall into three simple classes. Rational asymptotically free models
give $x^{-1}/\ln^{2}(1/x)$, diagonal-scattering models give $f(x)\propto x$,
and Sine-Gordon theory gives a coupling-dependent power $x^{-\gamma}$. We
tested these functions against the HERA data by allowing an independent
normalization at each $Q^{2}$. This is the fairest direct test of the $x$
dependence because the exact $(1+1)$-dimensional factor $g(Q^{2})$ differs
from the full four-dimensional HERA normalization. The rational and diagonal
classes give very large reduced $\chi^{2}_{\nu}$ values. Among the leading
templates explicitly tested, the Sine-Gordon power-law form supplies a
compatible rising small-$x$ behavior, and the finite-$Q^{2}$ Sine-Gordon
crossover gives $\chi^{2}_{\nu}=1.166$ for explicit HERA $F_{2}$ and $\chi
^{2}_{\nu}=1.172$ for the final HERA $\sigma_{r}$ proxy.

\begin{figure}[t]
\centering
\includegraphics[width=\columnwidth]{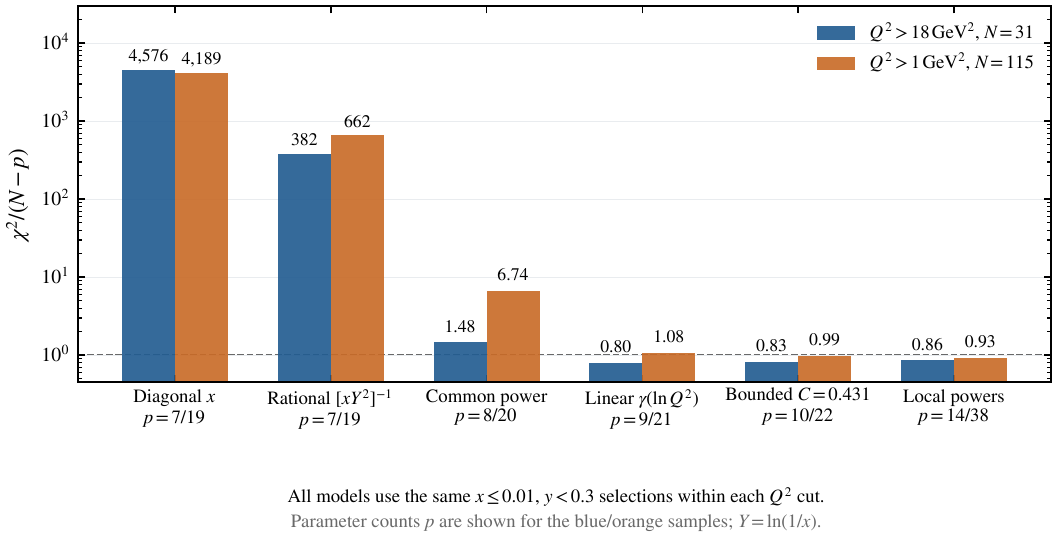}\caption{Direct
discriminator between the exact small-$x$ functions of the integrable models
and the HERA data. Independent normalizations are allowed at each $Q^{2}$ for
the first three asymptotic $x$ functions. The bounded crossover uses
independent slice normalizations, separately from the compact surface in
Eq.~\eqref{eq:best_sg_F2}. The hierarchy supports local powers and
compatibility with the reference crossover, without a unique
universality-class determination.}%
\label{fig:exact_function_comparison}%
\end{figure}

The positive finite-window exponent is compatible with the Sine-Gordon
reference $C=0.431$. Equation~\eqref{beataQ} converts the measured slope into
an effective coupling conditional on the proposed longitudinal description.
The profile range $0.298<C<0.483$ inherits the kinematic cuts and crossover
assumptions; it is not an independent measurement of a QCD coupling.

Matching the illustrative reference exponent $C=0.431$ in
Eq.~\eqref{eq:gamma_fit} to Eq.~\eqref{eq:sg_gamma} gives
\[
C=3-\frac{16\pi}{\beta^{2}}, \qquad\beta^{2}=\frac{16\pi}{3-C}, \qquad
\frac{\beta^{2}}{\pi}\simeq6.23.
\]
This gives $\nu=2/(1-C)=3.515$ and $\gamma\simeq0.431$. The matched coupling
lies safely in the repulsive solitonic window $4<\beta^{2}/\pi<8$. Using the
profile interval $0.298<C<0.483$ gives $5.922<\beta^{2}/\pi<6.357$, while the
Sine-Gordon selected value remains $\beta^{2}/\pi=6.228$.

Exact integrable-theory results provide the nonperturbative templates; the
data determine which templates describe the measured shape. For Sine-Gordon,
exponent matching specifies the effective coupling at each $Q^{2}$.
Observables beyond that matching condition are needed to discriminate the
proposed dynamics.

The correspondence is therefore a quantitative candidate description, not a
uniqueness claim. The solitonic Sine-Gordon theory supplies an explicit
realization of the compatible power law.

The match is not unique at the level of one exponent. The exact-theory
analysis itself gives a formal $SL_{q}(N)$ generalization with $f(x)=4\sin
^{2}(\pi\zeta)x^{N\zeta-1}$~\cite{Babujian2025}. Here $N$ labels the algebra
and $\zeta$ denotes its deformation parameter, distinct from the Sine-Gordon
mass parameter $\mu$ in Eq.~\eqref{SG1}. A given $C$ fixes $N\zeta=1-C$ but
does not determine $N$. This observation neither derives a local effective QCD
theory for every member nor selects Sine-Gordon uniquely. Within the examples
explicitly compared here, Sine-Gordon is the concrete solitonic model that
supplies the compatible power-law form.

The fixed-coupling normalization provides a separate exact-model calculation
within the chosen spectral truncation. We calculate the
three-intermediate-particle contribution $g_{3}(z)$, $z=Q^{2}/M^{2}$, from the
published current form factors rather than extrapolating the phenomenological
polynomial. At $C=0.431$ we obtain
\begin{equation}
g_{3}(18)=0.16332508,\qquad g_{3}(1000)=0.51673857,\qquad g_{3}(\infty
)=0.589320245.\label{eq:g3_results}%
\end{equation}
The numerical convergence and an independent integration are documented in
appendix~\ref{app:g3}. Positive spectral weight guarantees a monotone
increase, and its large-rapidity tail gives
\begin{equation}
g_{3}(\infty)-g_{3}(z)\propto z^{-p},\qquad p=\frac{\nu+1}{2\nu}=\frac{3-C}%
{4}=0.64225.\label{eq:g3_tail}%
\end{equation}
This exponent concerns the normalization in the three-particle approximation.
It is not the crossover parameter $\omega$ in Eq.~\eqref{eq:sg_crossover_F2}.

\begin{figure}[t]
\includegraphics[width=\columnwidth]{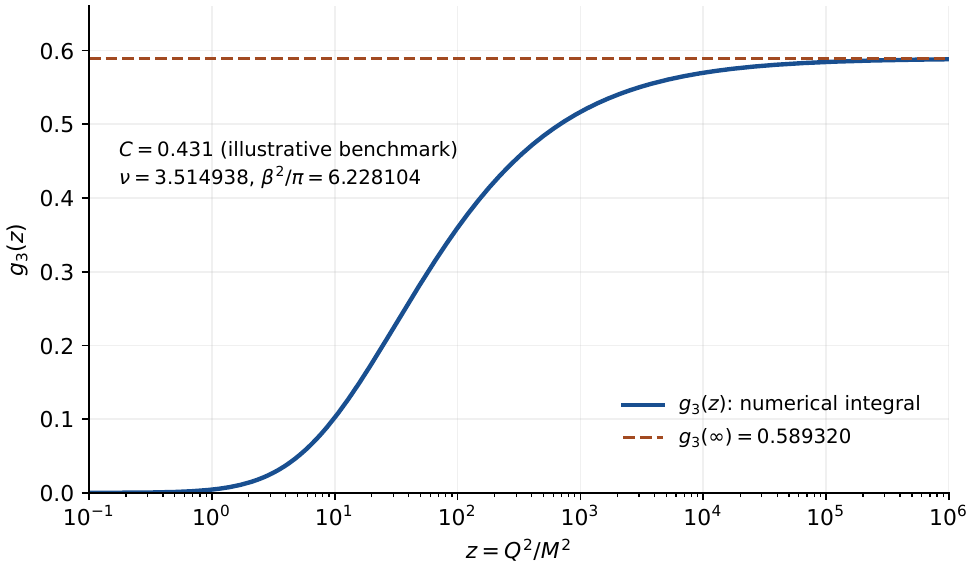}\caption{Three-particle
Sine-Gordon normalization $g_{3}(z)$ at the fixed benchmark $C=0.431$,
$\nu=3.514938489$, $z=Q^{2}/M^{2}$. The independently integrated asymptote is
$0.589320245$. No identification of $M$ with the proton mass is made.
Numerical integration checks do not bound omitted higher-particle
contributions or establish the physical DIS normalization.}%
\label{fig:sgnorm}%
\end{figure}

The normalization identifies a further test of the proposed correspondence. At
a fixed Sine-Gordon coupling the full prefactor is $4\sin^{2}(\pi/\nu
)g(Q^{2}/M^{2})$, and the scale dependence resides in $g$.
Figure~\ref{fig:sgnorm} shows the three-particle approximation at the fixed
benchmark coupling. With the scale-dependent prescription of
Eq.~\eqref{beataQ}, both $\nu$ and the form factors would also depend on the
matched coupling. The fixed-coupling plot therefore illustrates the exact
model, rather than supplying a parameter-free normalization prediction for the
running-coupling proposal. Such a test requires the mass scale and
electromagnetic-current matching.

The resulting picture is a hard current preparing a high-energy channel whose
subsequent longitudinal evolution is represented by collective world-sheet
dynamics. The measured exponent specifies the coupling prescription, while an
eventual QCD derivation must determine the field, current, and mass-scale matching.

The remaining derivation must identify the QCD configurations represented by
the world-sheet scalar and its soliton--antisoliton excitations, fix the
electromagnetic-current operator and hard coefficient, and relate $Q$ to
world-sheet scale evolution. These steps would turn the exponent
correspondence into predictions for additional DIS observables.

\section{Summary}

\label{sec:summary}

We have tested a local-power description of small-$x$ DIS using the published
correlated HERA uncertainties and compared its exponent with exact
integrable-model results. The primary data support $a(Q^{2})x^{-\gamma(Q^{2}%
)}$ over the specified window, while the strict product form $a(Q^{2}%
)\mathcal{F}(x)$ fails across the full selected scale range. This distinction
separates measured scaling violations from a proposed asymptotic factorization.

The final HERA check makes the domain of this statement precise. In the final
HERA I+II reduced-cross-section data with $Q^{2}>1\,\mathrm{GeV}^{2}$,
$x\le0.01$, $y<0.3$, the strict product benchmark gives $\chi^{2}_{\nu
}=230.373/80=2.880$, a common fixed power gives $\chi^{2}_{\nu}%
=640.096/95=6.738$, and the local-power form gives $\chi^{2}_{\nu
}=71.811/77=0.933$. The explicit HERA $F_{2}$ table gives the same conclusion:
the compact Sine-Gordon-constrained crossover fit gives $\chi^{2}_{\nu
}=123.608/106=1.166$. Thus the local-power description is a quantitative
property of the data and the Sine-Gordon plateau gives a compatible physical
interpretation of the exponent.

The measured exponent rises with $Q^{2}$ over the enlarged range. The
final-HERA slopes at $Q^{2}=60$ and $90\,\mathrm{GeV}^{2}$ are $0.299\pm0.008$
and $0.304\pm0.015$. The finite-$Q^{2}$ approach to the reference plateau is
described by the bounded crossover in Eq.~\eqref{eq:sg_crossover_F2}. This
explains why the data look nearly linear on a logarithmic $Q^{2}$ axis over
the HERA interval, while remaining compatible with a finite asymptotic exponent.

The main proposal is a Sine-Gordon description of the longitudinal DIS sector
on an effective string world sheet, with its scale-dependent coupling fixed by
exponent matching. In the analyzed window $Q^{2}>1\,\mathrm{GeV}^{2}$,
$x\le0.01$, $y<0.3$, we parametrize the structure function $F_{2}(x,Q^{2})$ by
the form
\begin{equation}
F_{2}(x,Q^{2})=\mathcal{F}_{2}\!\left( x^{-\gamma(Q^{2})}\right) a(Q^{2}),
\end{equation}
resembling the Regge-type factorization in Eq.~(\Ref{regge1}). Moreover, at
sufficiently large $Q^{2}$, the Sine-Gordon theory with
\begin{equation}
\label{beataQ}\beta(Q^{2})=\sqrt{\frac{16\pi}{3-\gamma(Q^{2})}}%
\end{equation}
is proposed to encode the longitudinal small-$x$ dynamics.
Equation~\eqref{beataQ} defines an effective coupling at the probe scale; its
scale dependence is not independently derived as a world-sheet beta function.

The fitted $\gamma(Q^{2})$ rises from about $0.15$ at $Q^{2}=1.5\,\mathrm{GeV}%
^{2}$ to about $0.30$ at $Q^{2}=90\,\mathrm{GeV}^{2}$. The bounded
extrapolation is compatible with the reference limiting value $\gamma
(\infty)=0.431$. This number is the conditional bridge from HERA data to exact
two-dimensional theory. The associated quartic normalization is an
interpolation on the fitted range and has no claimed asymptotic value. A
string-inspired interpretation motivates this longitudinal description: after
short-distance transverse physics has been matched, a two-dimensional world
sheet may represent the remaining collective high-energy channel. We assume
that a scalar field with a periodic interaction provides its leading dynamics.
The suppression or integration of transverse modes, and the identification of
this scalar in QCD, require a separate derivation. The proposed world-sheet
dynamics is expressed schematically by the action below, with the coupling
prescribed at each probe scale by exponent matching.

In the repulsive regime $\beta^{2}>4\pi$, the spectrum is the
soliton--antisoliton sector. A soliton is a topological kink of the field, and
an antisoliton is the kink with opposite topological charge. In that regime
the exact form-factor analysis gives~\cite{Babujian2025}%
\begin{equation}
S_{\mathrm{MT}}[\varphi;Q^{2}]=\int d^{2}\xi\,\left[  \bar{\psi}%
(i\gamma\partial-M)\psi-\frac{1}{2}g(Q^{2})(\bar{\psi}\gamma^{\mu}\psi
)^{2}\right]  ,\label{SG2}%
\end{equation}
where $g(Q^{2})=-\frac{1}{4}\pi\left(  1+\gamma(Q^{2})\right)  $ is given by use of Eq.~\eqref{beataQ}

 We use the canonical Minkowski field normalization for the
exponent--coupling relation. At each fixed probe scale, $\beta(Q^{2})$ is a
parameter independent of the world-sheet coordinates, and varying this action
gives Eq.~\eqref{SG1}. Its $Q^{2}$ dependence labels the matching prescription
rather than an extra world-sheet coordinate.

Exact form-factor results in $(1+1)$-dimensional integrable models then give a
clean discriminator. Rational asymptotically free models predict the
Balog--Weisz small-$x$ form
\[
F_{2}\sim\frac{1}{x\ln^{2}(1/x)},
\]
whereas diagonal-scattering models give $f(x)\propto x$. The observed HERA
hierarchy selects the remaining power-law class. The Sine-Gordon model gives
\[
F_{2}\sim x^{-\gamma}, \qquad\gamma=3-\frac{16\pi}{\beta^{2}},
\]
in the repulsive solitonic regime. At the illustrative limiting benchmark, the
conditional matching gives
\[
C=0.431, \qquad\nu=3.515, \qquad\frac{\beta^{2}}{\pi}=6.228 .
\]
The broader data-driven profile interval $0.298<C<0.483$ maps to
$5.922<\beta^{2}/\pi<6.357$, which remains inside the same solitonic
Sine-Gordon regime.

The three-particle normalization calculation adds a quantitative benchmark
within the candidate model. Starting from the current form factors of
Ref.~\cite{Babujian2025}, we obtain $g_{3}(\infty)=0.589320245$ at the
reference coupling, check the integration independently, and derive its
monotonicity and asymptotic approach. Appendix~\ref{app:g3} gives the
calculation and its truncation limits; appendix~\ref{app:statistics} documents
the covariance treatment, cut dependence, and broader-window comparison.

The proposed reduction makes the longitudinal small-$x$ behavior accessible to
nonperturbative integrable methods. The empirical exponent motivates
Sine-Gordon as a concrete solitonic realization and fixes its scale-dependent
matching prescription.

The HERA results and exact-model calculation provide quantitative support for
investigating a Sine-Gordon world-sheet description of the longitudinal DIS
sector. The next test is to determine the QCD current and scale matching and
confront an observable beyond the fitted exponent. This would establish the
predictive scope of the proposed correspondence.

\acknowledgments
The authors thank T.~Sedrakyan, A.~Tsvelik, L.~Radzihovsky, G.~Karyan,
S.~Derkachev and A.~Sherman for valuable discussions. H.B. and A.S.
acknowledge support from the Armenian HESC grants 21AG-1C024 (H.B., A.S.) and
24FP-1F039 (A.S.). M.K. thanks Felix von Oppen for hospitality at the Institut
f\"ur Theoretische Physik, Freie Universit\"at Berlin. Part of the research
reported in this paper was performed at the Max-Planck-Institut f\"ur Physik
komplexer Systeme, whose support is gratefully acknowledged.

\appendix

\section{Reproducible statistical analysis and cut dependence}

\label{app:statistics}

\subsection{Covariance and exact elimination of normalizations}


Let $X_{is}=1$ if point $i$ belongs to virtuality slice $s$, and zero
otherwise, and let $H_{is}=X_{is}\ln(1/x_{i})$. Then $m=Xa+H\gamma$, with
$a_{s}=\ln a(Q_{s}^{2})$. For $A=(X,H)$ the generalized least-squares solution
is
\begin{equation}
\widehat b=(A^{T}V^{-1}A)^{-1}A^{T}V^{-1}z,\qquad\operatorname{Cov}(\widehat
b)=(A^{T}V^{-1}A)^{-1}.\label{eq:gls_solution}%
\end{equation}
The slope block gives $\widehat\gamma$ and its full covariance $V_{\gamma}$.
For any specified crossover $\gamma(Q_{s}^{2};\theta)$, eliminating the slice
normalizations gives the exact quadratic identity
\begin{equation}
\min_{a}\chi^{2}[a,\gamma(\theta)] =\chi^{2}_{\mathrm{local}} +[\widehat
\gamma-\gamma(\theta)]^{T}V_{\gamma}^{-1} [\widehat\gamma-\gamma
(\theta)].\label{eq:sufficient_slopes}%
\end{equation}
This allows exponent fitting without discarding cross-slice correlations.
Independent direct nonlinear fits to all data reproduce the compressed
calculation within $10^{-9}$ in $\chi^{2}$ for the primary and broader samples
at both thresholds. A separate fit with $169$ Gaussian-penalized nuisance
shifts gives the same result.

Specifically, writing $V=D+SS^{T}$, where $D$ is the uncorrelated diagonal
covariance, the nuisance formulation is
\begin{equation}
\chi^{2}=\min_{b}\left[ (r-Sb)^{T}D^{-1}(r-Sb)+b^{T}b\right] ,\qquad
b=(I+S^{T}D^{-1}S)^{-1}S^{T}D^{-1}r,\label{eq:nuisance}%
\end{equation}
with $r=z-m$. For the primary $Q^{2}>1$ fixed-$C$ crossover the pointwise term
is $83.7181$ and the systematic penalty is $8.0298$, summing to $91.7480$. The
largest profiled pointwise pull is $2.16$ and the largest source shift is
$0.864$ standard deviation.

The local model has $2K$ fitted parameters for $K$ slices; the logarithmic
exponent has $K+2$, the fixed-$C$ crossover $K+3$, and the free-$C$ crossover
$K+4$. The rational and diagonal templates each have $K$. The compact
quartic-normalization fit has eight parameters when $C$ is fixed. All
comparisons on a given sample use the same covariance. The displayed
goodness-of-fit probabilities and conditional parameter errors refer to this
fixed log-Gaussian model. They do not include an uncertainty for the assumed
$F_{2}$ conversion or the extrapolation form.

\subsection{Broad-window high-virtuality check}

The full broad-window refit retains the high-$Q^{2}$ observations while
lowering the threshold to $Q^{2}>1\,\mathrm{GeV}^{2}$. The selection
$x\le0.18$, $y<0.6$, with at least three distinct $x$ points per slice, gives
294 final-HERA and 313 HERA-I $F_{2}$ observations, each in 33 slices over
$1.2\le Q^{2}/\mathrm{GeV}^{2}\le3000$. The two samples overlap and are fitted
separately. With independent slice normalizations, the fixed-$C=0.431$
crossover gives $378.394/258=1.467$ for final HERA and $341.911/277=1.234$ for
$F_{2}$, with probabilities $1.47\times10^{-6}$ and $0.00472$. The compact
quartic-normalization surfaces give $439.989/286=1.538$ and
$406.401/305=1.332$, respectively. Thus the broader window retains significant
residual disagreement; it is not used to claim a satisfactory global
description. Its higher-scale slices probe larger $x$ than the primary
selection and do not measure the same restricted small-$x$ slope. The
corresponding free-$C$ independent-normalization minima are $0.4627$ and
$0.4643$, demonstrating the dependence of the extrapolated endpoint on the
selected window. Machine-readable broad-window results accompany the analysis.

\subsection{Measured exponents and stability}

Table~\ref{tab:slopes} gives the joint final-data exponents for the primary
$Q^{2}>1$ sample. These are marginal values from the fit to all selected
slices. Refitting only $Q^{2}>18$ changes them slightly because the
correlated-source constraints change; both selected samples and their full
covariance matrices are provided separately.

\input{primary_slopes.tex} \begin{table}[t]
\centering{\small
\begin{tabular}
[c]{lrrrr}%
\toprule Additional selection ($x\le0.01$) & $N$ & $K$ & Local power &
Crossover, $C=0.431$\\
\midrule $Q^{2}>18$, $y<0.2$ & 19 & 5 & $8.65/9$ & $9.09/11$\\
$Q^{2}>1$, $y<0.2$ & 93 & 17 & $54.22/59$ & $73.42/73$\\
$Q^{2}>1$, $y<0.25$ & 104 & 18 & $65.83/68$ & \textemdash\\
$Q^{2}>3.5$, $y<0.3$ & 76 & 14 & $44.87/48$ & $52.64/59$\\
$Q^{2}>18$, $y<0.6$ & 61 & 9 & $58.25/43$ & $60.68/49$\\
$Q^{2}>1$, $y<0.6$ & 163 & 21 & $180.21/121$ & $199.25/139$\\
\bottomrule &  &  &  &
\end{tabular}
}\caption{Final-data cut variations with full covariance. Virtualities are in
$\mathrm{GeV}^{2}$ and all lower thresholds are strict. Entries are $\chi
^{2}/\mathrm{dof}$. The $y<0.25$ row is a local-power-only diagnostic.}%
\label{tab:cut_stability}%
\end{table}

The profile calculation refits the nonlinear parameters with multiple starting
values. Its numerical ranges are $-2\le\gamma_{\mathrm{soft}}\le0.6$,
$-8\le\ln(Q_{*}^{2}/\mathrm{GeV}^{2})\le35$, and $0.005\le\omega\le50$, with
$\gamma_{\mathrm{soft}}<C$ at fixed $C$. The primary high-cut minimum is
interior, at $\omega=7.83$. Widening the remaining bounds, or imposing
$\gamma_{\mathrm{soft}}\ge0$, does not remove the high-cut lack of an upper
limit on $C$. Some broader-window minima meet bounds; the $Q^{2}>18$, $y<0.2$
diagnostic also reaches the $\omega=50$ ceiling. No Gaussian
asymptotic-parameter errors are inferred from those cases. The lower-cut
primary minimum is interior, but the cut variations in the main text remain
part of its interpretation.

The origin of the extrapolation ambiguity is also analytic. Near the midpoint
$t=\ln(Q^{2}/Q_{*}^{2})=0$,
\begin{equation}
\gamma(t)=\frac{\gamma_{\mathrm{soft}}+C}{2} +\frac{\omega(C-\gamma
_{\mathrm{soft}})}{4}t -\frac{\omega^{3}(C-\gamma_{\mathrm{soft}})}{48}%
t^{3}+O(t^{5}).\label{eq:crossover_degeneracy}%
\end{equation}
The same midpoint $\gamma_{0}$ and local slope $b$ are reproduced by the
family $\gamma_{\mathrm{soft}}=2\gamma_{0}-C$, $\omega=2b/(C-\gamma_{0})$. A
nearly linear measured segment therefore cannot identify a unique $C$.
Curvature or a longer independent high-scale lever arm is needed.

\FloatBarrier

\section{Three-particle Sine-Gordon normalization}

\label{app:g3}
We use the current convention and three-intermediate-particle spectral
integral of Ref.~\cite{Babujian2025}, Eqs.~(5.2), (5.11) and (D.4). With
$z=Q^{2}/M^{2}$ and real relative rapidity $u$, define
\begin{align}
K_{\nu}(t) & =\frac{\sinh[(1-\nu)t/2]}{\sinh(\nu t/2)\cosh(t/2)}%
,\label{eq:ff_kernel}\\
\ln|F(u)|^{2} & =\int_{0}^{\infty}\frac{dt}{t}K_{\nu}(t) \frac{1-\cosh
t\cos(ut/\pi)}{\sinh t},\label{eq:ff_norm}\\
w_{\nu}(u) & =(1+e^{-2u/\nu}) \frac{2\tanh^{2}(u/2)}{\cosh(u/\nu)+\cos(\pi
/\nu)}|F(u)|^{2},\label{eq:spectral_weight}\\
g_{3}(z) & =\frac{1}{2\pi}\int_{0}^{\infty}\frac{w_{\nu}(u)\,du}{[1+(2+2\cosh
u)/z]^{2}}.\label{eq:g3_integral}%
\end{align}
The apparent cancellation at $t=0$ is evaluated using
\begin{equation}
\frac{1-\cosh t\cos(ut/\pi)}{\sinh t} =2\coth t\sin^{2}\!\frac{ut}{2\pi}%
-\tanh(t/2),
\end{equation}
whose ratio to $t$ tends to $[(u/\pi)^{2}-1]/2$. Exponential forms of the
hyperbolic functions avoid overflow at large integration variables.

\begin{table}[t]
\centering
\begin{tabular}
[c]{rr@{\qquad}rr}%
\toprule $z$ & $g_{3}(z)$ & $z$ & $g_{3}(z)$\\
\midrule 1 & 0.004475711 & 1000 & 0.516738571\\
10 & 0.102083581 & 2000 & 0.539901306\\
18 & 0.163325077 & 10000 & 0.569811757\\
100 & 0.359564365 & $\infty$ & 0.589320245\\
\bottomrule &  &  &
\end{tabular}
\caption{Three-particle normalization at $C=0.431$, $\nu=3.514938489$. The
argument $z=Q^{2}/M^{2}$ is dimensionless.}%
\label{tab:g3}%
\end{table}

Nested adaptive quadrature with $(u_{\max},t_{\max})=(30,35)$ and $(45,55)$
changes $g_{3}(\infty)$ by $1.05\times10^{-8}$; increasing $u_{\max}$ to $50$
changes it by less than $7\times10^{-13}$. An independent $1024$-node
Gauss--Legendre integration in $t$, using the unsimplified integrand, followed
by a $9001$-node Simpson integration in $u$, agrees within $1.2\times10^{-12}$
in absolute value at the checked arguments. Table~\ref{tab:g3} quotes only
digits supported by these checks. They constrain numerical integration errors
within $g_{3}$, not the omitted five- and higher-particle contributions to $g$.

There are two useful analytic checks. First, $w_{\nu}(u)\ge0$ and
$h(u)=2+2\cosh u>0$, so
\begin{equation}
\frac{d g_{3}}{d\ln z} =\frac{1}{\pi}\int_{0}^{\infty}du\,w_{\nu}(u)
\frac{h(u)/z}{[1+h(u)/z]^{3}}>0.
\end{equation}
Second, $K_{\nu}(0)=(1-\nu)/\nu$ implies $\ln|F(u)|^{2}=-(\nu-1)u/(2\nu)+O(1)$
at large $u$. Thus $w_{\nu}(u)\sim B e^{-pu}$ with $p=(\nu+1)/(2\nu
)\in(1/2,1)$. Substituting $v=e^{u}/z$ in the difference of the two spectral
integrals gives
\begin{equation}
g_{3}(\infty)-g_{3}(z)\sim\frac{B(1+p)}{2\sin(\pi p)}z^{-p}.
\end{equation}
This proves the approach stated in Eq.~\eqref{eq:g3_tail} within the specified
spectral truncation, independently of the phenomenological normalization fit
to HERA.

\subsection{Exact scattering-amplitude asymptotics}

As a further check of the power's analytic origin, the exact
soliton--antisoliton reflection amplitude satisfies $c(\theta)/a(\theta
)=\sinh(i\pi/\nu)/ \sinh[(i\pi-\theta)/\nu]$~\cite{Babujian2025}. Elastic
unitarity gives $|a(\theta)|=1$ for real $\theta$, hence
\begin{equation}
|c(\theta)|^{2}= \frac{\sin^{2}(\pi/\nu)}{\sinh^{2}(\theta/\nu)+\sin^{2}%
(\pi/\nu)} \sim4\sin^{2}(\pi/\nu)e^{-2\theta/\nu}.
\end{equation}
Writing $r=e^{-2\theta/\nu}$ and $s_{\nu}^{2}=\sin^{2}(\pi/\nu)$, the relative
error of the leading expression is exactly
\begin{equation}
\frac{4s_{\nu}^{2}r}{|c(\theta)|^{2}}-1=(4s_{\nu}^{2}-2)r+r^{2}.
\end{equation}
At the benchmark it falls below $10\%$, $5\%$ and $1\%$ for $\theta>3.14166$,
$4.13358$ and $6.69891$, respectively. This checks an exact scattering
building block. It is not a bound on finite-$x$ corrections to the complete
structure function, since the spectral integration and its kinematic mapping
remain to be controlled.

\end{document}

%% file: primary_slopes.tex
\begin{table}[t]
\centering\small
\begin{tabular}{rrrrrr}
\toprule
$Q^2\ (\mathrm{GeV}^2)$ & $N_\sigma$ & $\gamma(\sigma_r)$ & $\gamma(F_2)$ & $x_{\min}$ & $x_{\max}$\\
\midrule
1.2 & 7 & $0.1748\pm0.0109$ & $0.1763\pm0.0109$ & $6.3\times10^{-5}$ & $5\times10^{-3}$\\
1.5 & 9 & $0.1508\pm0.0085$ & $0.1528\pm0.0086$ & $5\times10^{-5}$ & $3.2\times10^{-3}$\\
2 & 7 & $0.1634\pm0.0076$ & $0.1649\pm0.0074$ & $8\times10^{-5}$ & $3.2\times10^{-3}$\\
2.7 & 8 & $0.1661\pm0.0059$ & $0.1638\pm0.0058$ & $1.3\times10^{-4}$ & $5\times10^{-3}$\\
3.5 & 8 & $0.1705\pm0.0037$ & $0.1726\pm0.0037$ & $1.3\times10^{-4}$ & $8\times10^{-3}$\\
4.5 & 7 & $0.1927\pm0.0057$ & $0.1944\pm0.0057$ & $2\times10^{-4}$ & $3.2\times10^{-3}$\\
6.5 & 7 & $0.2003\pm0.0053$ & $0.2021\pm0.0054$ & $3.2\times10^{-4}$ & $5\times10^{-3}$\\
8.5 & 7 & $0.2097\pm0.0060$ & $0.2145\pm0.0061$ & $3.2\times10^{-4}$ & $5\times10^{-3}$\\
10 & 5 & $0.2310\pm0.0104$ & $0.2330\pm0.0104$ & $5\times10^{-4}$ & $5\times10^{-3}$\\
12 & 6 & $0.2268\pm0.0071$ & $0.2316\pm0.0072$ & $5\times10^{-4}$ & $5\times10^{-3}$\\
15 & 7 & $0.2348\pm0.0038$ & $0.2396\pm0.0038$ & $5\times10^{-4}$ & $8\times10^{-3}$\\
18 & 6 & $0.2411\pm0.0046$ & $0.2461\pm0.0045$ & $8\times10^{-4}$ & $8\times10^{-3}$\\
\midrule
22 & 6 & $0.2594\pm0.0104$ & $0.2753\pm0.0133$ & $8\times10^{-4}$ & $8\times10^{-3}$\\
27 & 5 & $0.2717\pm0.0062$ & $0.2761\pm0.0062$ & $1.3\times10^{-3}$ & $8\times10^{-3}$\\
35 & 5 & $0.2713\pm0.0057$ & $0.2783\pm0.0059$ & $1.3\times10^{-3}$ & $8\times10^{-3}$\\
45 & 4 & $0.2782\pm0.0080$ & $0.2851\pm0.0085$ & $2\times10^{-3}$ & $8\times10^{-3}$\\
60 & 4 & $0.2993\pm0.0082$ & $0.3129\pm0.0090$ & $2\times10^{-3}$ & $8\times10^{-3}$\\
70 & 4 & $0.3122\pm0.0183$ & $0.3246\pm0.0267$ & $2.5\times10^{-3}$ & $8\times10^{-3}$\\
90 & 3 & $0.3044\pm0.0154$ & $0.3148\pm0.0163$ & $3.2\times10^{-3}$ & $8\times10^{-3}$\\
\bottomrule
\end{tabular}
\caption{Joint local-power slopes for $Q^2>1\,\mathrm{GeV}^2$, $x\le0.01$, $y<0.3$.
The rule above $22\,\mathrm{GeV}^2$ separates the rows satisfying the strict $Q^2>18$ cut.
$N_\sigma$ and the $x$ endpoints refer to final cross sections; the HERA-I comparison is fitted separately with its own covariance.
Errors are marginal standard deviations, not independent uncertainties between rows or between the two data sets.}
\label{tab:slopes}
\end{table}